\newtheorem{dfn}{Definition}
\newtheorem{thm}[dfn]{Theorem}
\newtheorem{lmma}[dfn]{Lemma}
\newtheorem{ppsn}[dfn]{Proposition}
\newtheorem{crlre}[dfn]{Corollary}
\newtheorem{xmpl}[dfn]{Example}
\newtheorem{rmrk}[dfn]{Remark}

\newcommand{\bdfn}{\begin{dfn}}
\newcommand{\bthm}{\begin{thm}}
\newcommand{\blmma}{\begin{lmma}}
\newcommand{\bppsn}{\begin{ppsn}}
\newcommand{\bcrlre}{\begin{crlre}}
\newcommand{\bxmpl}{\begin{xmpl}}
\newcommand{\brmrk}{\begin{rmrk}}

\newcommand{\edfn}{\end{dfn}}
\newcommand{\ethm}{\end{thm}}
\newcommand{\elmma}{\end{lmma}}
\newcommand{\eppsn}{\end{ppsn}}
\newcommand{\ecrlre}{\end{crlre}}
\newcommand{\exmpl}{\end{xmpl}}
\newcommand{\ermrk}{\end{rmrk}}

\def\be{\begin{equation}}
\def\ee{\end{equation}}
\def\ba{\begin{array}}
\def\ea{\end{array}}

\def\Cb{{\Bbb C}}

\documentstyle[12pt]{article}
\begin{document}
\input amssym.def
\parskip=4pt
\parindent=18pt
\baselineskip=18pt \setcounter{page}{1}
\centerline{\Large\bf
Local Equivalence of Rank-Two Quantum Mixed States}
\vspace{4ex}
\begin{center}
Sergio Albeverio$^{a}$ \footnote{ SFB 611; BiBoS; IZKS; CERFIM(Locarno);
Acc. Arch.; USI(Mendrisio)}, Shao-Ming Fei$^{b,c}$, Debashish Goswami$^{d}$
\footnote{research partially supported by Av Humboldt Foundation}

\vspace{3ex}
\begin{center}
\begin{minipage}{5.6in}

{\small $~^{a}$ Institut f\"ur Angewandte Mathematik,
Universit\"at Bonn, D-53115}

{\small $~^{b}$ Department of Mathematics, Capital Normal
University, Beijing 100037}

{\small $~^{c}$ Max Planck Institute for Mathematics in the Sciences, D-04103 Leipzig}

{\small $~^{d}$ Statistics and Mathematics Unit, Indian Statistical
Institute, Kolkata 700108}

\end{minipage}
\end{center}
\end{center}

\vskip 1 true cm
\parindent=18pt
\parskip=6pt
\begin{center}
\begin{minipage}{5in}
\vspace{2ex} \centerline{\large Abstract} \vspace{4ex}

We investigate the equivalence of quantum mixed states under local
unitary transformations. For a class of rank-two mixed states,
a sufficient and necessary condition of local equivalence
is obtained by giving a complete set of invariants under local
unitary transformations, such that
two states in this class are locally equivalent
if and only if all these invariants have equal values for them.
\bigskip
\medskip
\bigskip
\medskip

PACS numbers: 03.67.-a, 02.20.Hj, 03.65.-w\vfill

\end{minipage}
\end{center}

Quantum entanglement has been extensively investigated
as a key physical resource to realize
quantum information tasks such as quantum cryptography, quantum
teleportation and quantum computation \cite{nielsen}.
Due the fact that the properties of entanglement for multipartite quantum systems
remain invariant under local unitary transformations on the subsystems,
the entanglement can be characterized in principle by all the invariants
under local unitary transformations. For instance,
the trace norms of realigned or partial transposed density
matrices in entanglement measure and separability criteria
are some of these invariants \cite{norm}.
Therefore a complete set of invariants
gives rise to the classification of the quantum states under local
unitary transformations. Two quantum states are locally
equivalent if and only if all these invariants have equal
values for these states.

There have been many results on calculation of invariants \cite{Rains,Grassl}
related to the equivalence of quantum states under local unitary transformations,
e.g. for general two-qubit systems \cite{Makhlin},
three-qubit states \cite{Linden99,sun3qubit},
some generic mixed states \cite{generic,goswami,sungeneric}, some classes of
tripartite pure and mixed states \cite{wl}. However till now we still
have no operational criteria to judge the equivalence for two
general bipartite mixed states under local unitary transformations.
In this letter we investigate the local equivalence under
local unitary transformations for a class of
rank-two bipartite mixed quantum states in arbitrary dimensions,
and present an operational criterion.

Let $H_1$ and $H_2$ be $m$ and $n$-dimensional complex Hilbert spaces,
with $\vert e_\alpha\rangle$, $\alpha=1,...,m$,
and $\vert f_\beta\rangle$, $\beta=1,...,n$, $m\leq n$,
as orthonormal bases. Let $\rho_1$ and $\rho_2$ be two bipartite density
matrices defined on $H_1\otimes H_2$ with rank $r(\rho_1)=r(\rho_2)=2$.
$\rho_1$ and $\rho_2$ are said to be equivalent
under local unitary transformations if there exist unitary operators
$U_1$ on $H_1$ and $U_2$ on $H_2$ such that
\be\label{eq}
\rho_2=(U_1\otimes U_2)\rho_1(U_1\otimes U_2)^\dag,
\ee
where $\dag$ stands for transpose and conjugation.

As $\rho_1$ and $\rho_2$ are rank-two density matrices,
they have the following decompositions according to their eigenvalues and eigenvectors:
$$
\rho_i=\sum_{\alpha=1}^2\lambda_\alpha^i\vert\nu_\alpha^i\rangle\langle\nu_\alpha^i\vert,~~~i=1,2,
$$
where $\lambda_\alpha^i$ and $\vert\nu_\alpha^i\rangle$, $\alpha=1,2$, are the
nonzero eigenvalues and eigenvectors of the density matrix $\rho^i$ respectively,
$\sum_{\alpha=1}^2\lambda_\alpha^i=1$. $\vert\nu_\alpha^i\rangle$ has generally the form
$$
\vert\nu^i_1\rangle=\sum_{\alpha=1}^m\sum_{\beta=1}^n a_{\alpha\beta}^i \vert e_\alpha\rangle \otimes
\vert f_\beta\rangle,~~~
\vert\nu^i_2\rangle=\sum_{\alpha=1}^m\sum_{\beta=1}^n b_{\alpha\beta}^i \vert e_\alpha\rangle \otimes
\vert f_\beta\rangle,
$$
where $a_{\alpha\beta}^i,\,b_{\alpha\beta}^i\in\Cb$, $\sum_{\alpha\beta} a_{\alpha\beta}^i a_{\alpha\beta}^{i\ast}=
\sum_{\alpha\beta} b_{\alpha\beta}^i b_{\alpha\beta}^{i\ast}=1$, $i=1,2$, $\ast$ denotes complex conjugation.

Let $A_i$ and $B_i$ denote the $m\times n$ matrices with entries $a^{(i)}_{\alpha\beta}$
and $b^{(i)}_{\alpha\beta}$ respectively.
We consider the necessary and sufficient conditions of equivalence under local unitary
transformations for a
class of rank-two states satisfying the following conditions:
\be\label{cond}
A_i^\dag A_i=B_i^\dag B_i,~~A_iA_i^\dag=B_iB_i^\dag~~~{\rm  for}~ i=1,2.
\ee

{\sf [Theorem]} The density matrices $\rho_1$ and $\rho_2$
are equivalent under local unitary transformations if and
only if the following hold:

(i) $Tr(\rho_1^2)=Tr(\rho^2_2)$;

(ii) $Tr((A_1B_1^\dag )^\alpha)=Tr((A_2B_2^\dag )^\alpha)$,   $\forall$ $\alpha=1,...,m$;

(iii) $r(A_1)=r(A_2)$, $r(B_1)=r(B_2)$, $r((B_1^\dag A_1)^\alpha)=r((B_2^\dag A_2)^\alpha)$,  $\forall$ $\alpha=1,...,m$.

{[\sf Proof]} It is straightforward to see that (i)-(iii)
 above hold if $\rho_1$ and $\rho_2$ are equivalent under local
 unitary transformations, in the sense
 of eq.(\ref{eq}).

We prove the converse. Two pairs of ($m\times n$) matrices, $(A,B)$ and $(C,D)$, are called contragrediently equivalent
if $A = SCT^{-1}$, $B = TDS^{-1}$ for some invertible matrices $S$ and $T$.
It is shown in \cite{Holz} that the pairs $(A,B)$ and $(C,D)$ are contragrediently equivalent
if and only if $AB$ is similar to $CD$ and $r(A)=r(C)$, $r(B)=r(D)$,
$r(BA)^\alpha=r(DC)^\alpha$, $r(AB)^\alpha=r(CD)^\alpha$ for all $\alpha=1,...,m$.

Therefore from the conditions (ii) and (iii) we have that the pairs
 $(A_1,B_1^\dag )$ and $(A_2,B_2^\dag )$ are contragrediently equivalent and
 there are invertible (but not necessarily
 unitary) matrices $S$ and $T$ such that
\be\label{saat}
SA_2=A_1T,~~TB_2^{-1} =B_1^{-1} S.
\ee
 Eq. (\ref{saat}) can be rewritten as,
 $$
 \left( \begin{array}{cc} 0 & T \\ S & 0
 \end {array} \right)\left( \begin{array}{cc} 0 & A_2 \\ B_2^\dag  & 0
 \end{array}
 \right)=\left( \begin{array}{cc} 0 & B_1^\dag  \\ A_1 & 0
 \end {array} \right)\left( \begin{array}{cc} 0 & T \\ S & 0
 \end{array}
 \right).
 $$
 By assumption (\ref{cond}), the matrices $W_1:=\left(
 \begin{array}{cc} 0 & A_2 \\ B_2^\dag  & 0 \end{array} \right)$ and
 $W_2:=\left( \begin{array}{cc} 0 & B_1^\dag  \\ A_1 & 0 \end{array}
 \right)$ are normal.

If two normal matrices $M$, $N$ and an invertible matrix $X$ satisfy
$XMX^{-1}=N$, then one has $U_X M U_X^\dag =N$, where $X=U_X |X|$
is the polar decomposition of $X$ and $U_X$ is unitary \cite{rudin}.
Therefore from the observation that the
 unitary part of the polar decomposition in $\left(
 \begin{array}{cc} 0 & S\\ T & 0 \end{array} \right)$ is nothing but
 $\left( \begin{array}{cc} 0 & U_S\\ U_T & 0 \end{array} \right) $,
 we have
 $$
 \left( \begin{array}{cc} 0 & U_T \\ U_S & 0
 \end {array} \right)\left( \begin{array}{cc} 0 & A_2 \\ B_2^\dag  & 0
 \end{array}
 \right)=\left( \begin{array}{cc} 0 & B_1^\dag  \\ A_1 & 0
 \end {array} \right)\left( \begin{array}{cc} 0 & U_T \\ U_S & 0
 \end{array}
 \right),
 $$
 which is equivalent to
 $$
 A_2=U_S^\dag  A_1 U_T,~~B_2=U_S^\dag B_ 1 U_T.
 $$
Here $U_S$ and $U_T$ are unitary (as $S$, $T$ are invertible).
The condition (i) and $Tr(\rho_1)=Tr(\rho_2)=1$ together imply that the density matrices $\rho_1$
and $\rho_2$ have the same eigenvalues.
Therefore $\rho_2=(U_1\otimes U_2)\rho_1(U_1\otimes U_2)^\dag$,
where $U_1=U_S^\dag$, $U_2=(U_T)^t$ ($t$ denoting transpose).
\hfill $\Box$

The Theorem gives a sufficient and necessary condition for local equivalence
of two rank-two mixed states satisfying (\ref{cond}).
The class of quantum states satisfying (\ref{cond}) is not trivial.
As a simple example, we consider the two-qubit systems. In this case
$A$ and $B$ are $2\times 2$ matrices. It is easily verified that
the following matrices satisfy the required conditions,
$$
A(\theta)=\frac{1}{\sqrt{2}}\left( \begin{array}{cc} \cos\theta & \sin\theta \\ -\sin\theta & \cos\theta
 \end {array} \right)\,,~~~~~~
B(\gamma)=\frac{1}{\sqrt{2}}\left( \begin{array}{cc} \cos\gamma & \sin\gamma \\ \sin\gamma & -\cos\gamma
 \end {array} \right)\,.
$$
Hence the rank-two density matrix
$\rho=\lambda\vert\psi\rangle\langle\psi\vert + (1-\lambda)\vert\phi\rangle\langle\phi\vert$,
where
$\vert\psi\rangle=\sum_{\alpha,\beta=1}^2 a_{\alpha\beta}(\theta) \vert e_\alpha\rangle \otimes
\vert f_\beta\rangle$,
$\vert\phi\rangle=\sum_{\alpha,\beta=1}^2 b_{\alpha\beta}(\gamma) \vert e_\alpha\rangle \otimes
\vert f_\beta\rangle$, belongs to the class we are concerning.
From the theorem we have that $\rho$ and
$\rho^\prime=\lambda\vert\psi^\prime\rangle\langle\psi^\prime\vert
+ (1-\lambda)\vert\phi^\prime\rangle\langle\phi^\prime\vert$ with
$\vert\psi^\prime\rangle=\sum_{\alpha,\beta=1}^2 a_{\alpha\beta}(\theta^\prime) \vert e_\alpha\rangle \otimes
\vert f_\beta\rangle$,
$\vert\phi^\prime\rangle=\sum_{\alpha,\beta=1}^2 b_{\alpha\beta}(\gamma^\prime) \vert e_\alpha\rangle \otimes
\vert f_\beta\rangle$, are equivalent under local unitary transformations.

Here the concurrence $C(|\psi\rangle)=C(|\phi\rangle)=1$. Both pure states $|\psi\rangle$ and $|\phi\rangle$ are maximally entangled.
In the special case $\theta=0$ (resp. $\gamma=0$), $|\psi\rangle$ (resp. $|\phi\rangle$) is reduced to one of the
Bell bases $|\psi\rangle=(|00\rangle+|11\rangle)/\sqrt{2}$ (resp.
$|\phi\rangle=(|00\rangle-|11\rangle)/\sqrt{2}$). These states are
equivalent under local unitary transformations. Nevertheless, generally
$\rho$ and $\rho^\prime$ are not equivalent under local unitary transformations
even if $|\psi\rangle$ (resp. $|\phi\rangle$) is equivalent to $|\psi^\prime\rangle$ (resp. $|\phi^\prime\rangle$)
under local unitary transformations,
unless the same local unitary transformations transform
$|\psi\rangle$ to $|\psi^\prime\rangle$  and $|\phi\rangle$ to $|\phi^\prime\rangle$ simultaneously.

Generally a rank-two state can be written as
$\rho=\lambda\vert\nu_1\rangle\langle\nu_1\vert + (1-\lambda)\vert\nu_2\rangle\langle\nu_2\vert$,
$0<\lambda<1$. The normalized vectors $\vert\nu_1\rangle$ and $\vert\nu_2\rangle$
are given by the $m\times n$ matrices $(A)_{\alpha\beta}=a_{\alpha\beta}$ and
$(B)_{\alpha\beta}=b_{\alpha\beta}$ respectively,
$\vert\nu_1\rangle=\sum_{\alpha\beta} a_{\alpha\beta} \vert e_\alpha\rangle \otimes
\vert f_\beta\rangle$,
$\vert\nu_2\rangle=\sum_{\alpha\beta} b_{\alpha\beta} \vert e_\alpha\rangle \otimes
\vert f_\beta\rangle$, with $Tr(AA^\dag)=Tr(BB^\dag)=1$ due to normalization.
Let us consider the general forms of a pair of matrices $A$ and $B$ such that the conditions
$A^\dag A=BB^\dag$ and $AA^\dag=BB^\dag$ are satisfied.

Since  $A^\dag A=B^\dag B$,  we can write down  singular value decomposition of $A$ and $B$ as follows:
$$ A=U\Delta V^\dag, ~~B=U^\prime\Delta V^{\prime\dag},$$
where $U$, $U^\prime$ and $V$, $V^\prime$ are unitary matrices
and $\Delta$ is a diagonal matrix with nonnegative entries.
Furthermore, the condition $A^\dag A=B^\dag B$ implies  $V^{\prime\dag}V\Delta^2=\Delta^2 V^{\prime\dag}V$.
Thus,  $V^{\prime\dag}V$ commutes with $\Delta$.
Similarly from $A A^\dag=B B^\dag$ we conclude  that
$U^{\prime\dag}U$  also commutes with $\Delta$.
Hence we have
\be\label{B}
B=U^{\prime}\Delta V^{\prime\dag}=UU^{\dag}U^{\prime}\Delta V^{\prime\dag}VV^{\dag}
=U\Gamma\Delta V^{\dag},
\ee
where $\Gamma=(U^{\prime\dag}U)^\dag V^{\prime\dag}V$ is unitary and commutes with $\Delta$.
Therefore the pair $(A,B)$ can be transformed  into the pair $(\Delta,\Gamma\Delta)$.
We call $(\Delta,\Gamma\Delta)$ the canonical form of the pair $(A,B)$. If the diagonal matrix $\Delta$
is of the form $diag(d_1,...,d_1,d_2,...,d_2,...,d_k,...,d_k)$, where $d_i$ is repeated with
multiplicity $m_i$, then $\Gamma$ must have the block diagonal form $diag(\Gamma_1,...,\Gamma_k)$,
where $\Gamma_i$, $i=1,...,k$, are $m_i \times m_i$ unitary matrices.

In fact, if we have another pair of matrices $A^\prime$ and $B^\prime$, associated with
the eigenvectors $\vert\nu_1^\prime\rangle$ and $\vert\nu_2^\prime\rangle$ of another
rank-two density matrix in the class considered, with canonical form
$(\Delta^{\prime},\Gamma^{\prime}\Delta^{\prime})$,
then $\vert\nu_1^\prime\rangle$, $\vert\nu_2^\prime\rangle$
and $\vert\nu_1\rangle, \vert\nu_2\rangle$ are equivalent
under local unitary transformations if and only if
$\Delta=\Delta^{\prime}$ and $w_i\Gamma_i w_i^\dag =\Gamma^\prime_i$
for some unitary matrix $w_i$, $i=1,...,k$.

Therefore under the local unitary transformation $\rho~\to~(U\otimes V^\ast) \rho (U\otimes V^\ast)^\dag$,
a rank-two mixed state in our class has the standard form:
$\rho=\lambda\vert\mu_1\rangle\langle\mu_1\vert + (1-\lambda)\vert\mu_2\rangle\langle\mu_2\vert$,
where $\vert\mu_1\rangle=\sum_{\alpha} d_{\alpha} \vert e_\alpha\rangle \otimes
\vert f_\alpha\rangle$,
$\vert\mu_2\rangle=\sum_{\alpha\beta} (\Gamma\Delta)_{\alpha\beta} \vert e_\alpha\rangle \otimes
\vert f_\beta\rangle$. In particular,
if all the singular values are distinct, then two such density matrices are equivalent
under local unitary transformation if and if they have exactly the same standard form.

We have investigated the equivalence under local unitary transformations
for a class of rank-two bipartite mixed quantum states. A
complete set of invariants has been presented such that any two of
these states are locally equivalent if
and only if all these invariants have equal values for these related density matrices.

Our method can be applied to
another classification of quantum states, defined by
local operations and classical communication (LOCC).
Two states have the same kind of entanglement if they can be obtained from each other by LOCC
with nonzero probability \cite{Bennett}. There have been many results for bipartite
and multipartite pure states for their equivalence under SLOCC \cite{Dur,Verstraete,sofar}.
In \cite{Dur} D\"ur et al showed that for pure three-qubit states there are six different
classes of entanglement under SLOCC. Verstraete et al considered the entanglement of
four-qubit case under SLOCC and concluded that there exist
nine families of states corresponding to nine different ways of entanglement
\cite{Verstraete}. Nevertheless for mixed states few is known yet.

Corresponding to pure states, we say
that $\rho_1$ and $\rho_2$ are equivalent under SLOCC
if there exist {\it invertible} (but not
necessarily unitary) matrices $P$ and $Q$ such that
\be\label{slocc}
\rho_2=(P \otimes Q) \rho_1 (P \otimes Q)^\dagger.
\ee

{\sf [Proposition]} The density matrices $\rho_1$ and $\rho_2$, with $B_1$ and $B_2$ nonsingular,
are equivalent under SLOCC  if the following hold:

(i) $Tr(\rho_1^2)=Tr(\rho^2_2)$;

(ii) $Tr((A_1B_1^{-1} )^\alpha)=Tr((A_2B_2^{-1} )^\alpha)$,   $\forall$ $\alpha=1,...,m$;

(iii) $r(A_1)=r(A_2)$, $r(B_1)=r(B_2)$, $r((B_1^{-1} A_1)^\alpha)=r((B_2^{-1} A_2)^\alpha)$,  $\forall$ $\alpha=1,...,m$.

{[\sf Proof]} From the conditions (ii) and (iii) we have that the pairs
$(A_1,B_1^{-1} )$ and $(A_2,B_2^{-1} )$ are contragrediently equivalent.
Hence there are invertible (but not necessarily
unitary) matrices $S$ and $T$ such that $SA_2=A_1T$, $TB_2^{-1} =B_1^{-1} S$.

That is, we have $A_2=S^{-1} A_1 T,~~ B_2=S^{-1}B_1 T$.
Accounting to the condition (i) which implies that the density matrices $\rho_1$
and $\rho_2$ have the same eigenvalues, the above relations give rise to
the equivalence of $\rho_1$ and $\rho_2$ under SLOCC.
\hfill $\Box$

The classification of quantum states under local operations is of significance
in quantum information processing. We have presented some criteria for
the equivalence of some bipartite mixed states in arbitrary dimensions. Our results can be
generalized to the case of multipartite states by considering
bipartite decompositions. In terms of the method used in \cite{wl},
our equivalence criteria for bipartite mixed states can be
also used to study the equivalence of tripartite pure states.

\end{document}